\documentclass{ifacconf}

\usepackage{graphicx}      
\usepackage{natbib}        
\usepackage{color}
\usepackage{amsmath,mathrsfs}
\usepackage{amsfonts}
\newtheorem{remark}{Remark}
\usepackage{epstopdf}    
\usepackage{algorithm}
\usepackage{balance} 
\usepackage{algorithmicx}
\usepackage{algpseudocode}
\usepackage{xfrac}
\usepackage{bm}
\usepackage{pgfplots}
\usepgflibrary{shapes}
\usepackage{tikz}
\usepackage{enumitem}
\usepackage{verbatim}
\usetikzlibrary{arrows,calc,patterns}
\tikzstyle{Rect}=[draw=gray,line width=0.001pt,preaction={clip, postaction={pattern=north east lines, pattern color=gray,line width=0.1pt}}]
\tikzset{
	>=stealth',
	help lines/.style={dashed, thick},
	axis/.style={<->},
	important line/.style={thick},
	connection/.style={thick, dotted},
}

\usepackage{eso-pic}
\AddToShipoutPictureBG*{%
	\AtPageUpperLeft{%
		\setlength\unitlength{1in}%
		\hspace*{\dimexpr0.5\paperwidth\relax}
		\makebox(0,-1.75)[c]{
			\begin{tabular}{c c}
				Rodrigo A. Gonz\'alez et al.,
				Identifying Lebesgue-sampled Continuous-time Impulse Response Models: A Kernel-based Approach. \\
				To appear in
				{\em 22nd IFAC World Congress},
				Yokohama, Japan, 2023
				uploaded to ArXiv \today \\
\end{tabular}}}}

\begin{document}
\begin{frontmatter}

\title{Identifying Lebesgue-sampled Continuous-time Impulse Response Models: A Kernel-based Approach\thanksref{footnoteinfo}}

\thanks[footnoteinfo]{This work is part of the research program VIDI with project number 15698, which is (partly) financed by The Netherlands Organization for Scientific Research (NWO).}

\author[tue]{Rodrigo A. Gonz\'alez} 
\author[tue]{Koen Tiels}
\author[tue,delft]{Tom Oomen}

\address[tue]{Control Systems Technology Section, Department of Mechanical Engineering, Eindhoven University of Technology, The Netherlands. (e-mails: r.a.gonzalez@tue.nl; k.tiels@tue.nl; t.a.e.oomen@tue.nl).}
\address[delft]{Delft Center for Systems and Control, Delft	University of Technology, The Netherlands}

\begin{abstract}                
Control applications are increasingly sampled non-equidistantly in time, including in motion control, networked control, resource-aware control, and event-triggered control. Some of these applications use measurement devices that sample equidistantly in the amplitude domain. The aim of this paper is to develop a non-parametric estimator of the impulse response of continuous-time systems based on such sampling strategy, known as Lebesgue-sampling. To this end, kernel methods are developed to formulate an algorithm that adequately takes into account the output intersample behavior, which ultimately leads to more accurate models and more efficient output sampling compared to the standard approach. The efficacy of this method is demonstrated through a mass-spring damper case study.
\end{abstract}


\end{frontmatter}

\section{Introduction}
In system identification and control, it is common to assume that the signals are sampled equidistantly in time, also known as Riemann sampling. In contrast, event-based sampling can lead to improvements in control performance, as well as in resource efficiency \citep{aastrom2003systems}. In particular, Lebesgue sampling is one of the most popular event-based sampling methods, and it consists of sampling a signal whenever it crosses fixed thresholds in the amplitude domain. This sampling can be found in incremental encoders \citep{merry2013optimal}, and also in networked control systems, where the goal is to reduce resource utilization without affecting throughput.

The Lebesgue sampling paradigm provides knowledge on what amplitude band the signals are located in at each instant of time, thus yielding a quantized measurement whenever the signal does not cross a threshold. A considerable effort has been devoted to the identification of systems based on quantized measurements. The maximum likelihood estimator based on the expectation-maximization algorithm (EM) has been derived for finite-impulse response (FIR) systems by \cite{godoy2011identification}, while \cite{chen2012impulse} developed a regularized FIR estimator for binary measurements. \cite{risuleo2019identification} studied an approximate maximum likelihood approach, and \cite{bottegal2017new} propose a kernel-based method for estimating FIR models. Other approaches have been pursued for the identification of ARX systems \citep{aguero2017based}.

Although there has been recent work on non-parametric identification for continuous-time systems using kernel methods that use non-equidistantly sampled data \citep{scandella2022kernel}, these works do not incorporate the intersample behavior information derived from a Lebesgue sampling framework (i.e., the lower and upper bounds on the unsampled output in between the time-stamps), which leads to performance detriment. Other work \citep{kawaguchi2016system} has considered continuous-time systems with Lebesgue-sampled outputs, although such results are only valid for parametric models. On the other hand, the approaches by \cite{chen2012impulse,bottegal2017new,risuleo2019identification} for identification with quantized data might be used for obtaining a non-parametric discrete-time model that can later be converted into continuous-time. However, this conversion is in many cases ill-defined and does not admit an arbitrary input intersample behavior, which drives the need for directly estimating a continuous-time system from the data \citep{garnier2014advantages}. 

The problem that is addressed in this paper is the estimation of non-parametric continuous-time models from Lebesgue-sampled output data. To this end, we seek estimators that can 1) provide a \textit{continuous-time} impulse response estimate from possibly noisy and short data records, and 2) exploit the entirety of the output information contained in the irregular sampling instants and the set-constrained intersample behavior. In summary, the main contributions of this paper are:
\begin{enumerate}[label=(C\arabic*)]
	\item We propose a non-parametric continuous-time impulse response estimator for Lebesgue-sampled systems. This estimator is obtained by first introducing a loss function that incorporates the output intersample information together with a regularization term. The optimum of the infinite-dimensional optimization problem is characterized by the representer theorem \citep{scholkopf2001generalized}.
	
	\item Once the kernel-regularized estimator is written as a finite linear combination of representers, we present an iterative procedure that delivers the associated weights based on the \textit{maximum a posteriori} EM (MAP-EM) method. This is done by relating the optimization problem for computing the weights to the MAP estimator of a particular FIR model. 
	
	\item We introduce an iterative scheme to optimize the hyperparameters that describe the kernel and noise variance from the Empirical Bayes (EB) approach. These iterations also stem from MAP-EM and require the computation of second-order moments of the unsampled output, which are approximated via a minimax tilting algorithm \citep{botev2017normal}.
\end{enumerate}

The remainder of the paper is organized as follows: in Section \ref{sec:problemformulation} the problem of interest is stated, and practical aspects of Lebesgue-sampled system identification are covered. Section \ref{sec:kernel} contains the derivation of a kernel-based estimator for continuous-time Lebesgue-sampled system identification. A numerical study is presented in Section \ref{sec:simulations}, while Section \ref{sec:conclusions} provides concluding remarks. Proofs of several theoretical results can be found in the Appendix.

\section{Problem formulation}
\label{sec:problemformulation}
Consider the following linear and time-invariant (LTI), stable, strictly causal, continuous-time system
\begin{equation}
	\label{system}
	x(t) = \int_{0}^\infty g(\tau) u(t-\tau)\textnormal{d} \tau,
\end{equation}
where $u$ is a causal, deterministic and exogenous input, and $g$ is the impulse response of the LTI system. The output $x(t)$ is corrupted by additive measurement noise, which results in a continuous-time signal $z(t)$. Assume that we have access to $N_{\textnormal{L}}$ data points of the Lebesgue-sampled version of $z(t)$, as in Fig. \ref{fig1}. That is, given the threshold distance $h>0$ and the signal $z(t)$, we have at our disposal the sampled sequence $\{y_\textnormal{L}(t_l)\}_{l=1}^{N_\textnormal{L}}$ that satisfies $y_\textnormal{L}(t_l)\hspace{-0.04cm}=\hspace{-0.04cm}z(t_l)$. The time-stamps $t_l, l=1,2,\dots,N_\textnormal{L}$, are the instants in time at which $z(t)$ crosses a fixed threshold $h m_l$, with $m_l\hspace{-0.04cm}\in\hspace{-0.04cm} \mathbb{Z}$. Formally, we characterize the time-stamps by
\begin{align}
	t_l = \min \left\{ \hspace{-0.03cm}\tau \in (t_{l-1},\infty): z(\tau) = mh \textnormal{ for some }m\in \mathbb{Z}\right\}, \notag
\end{align}
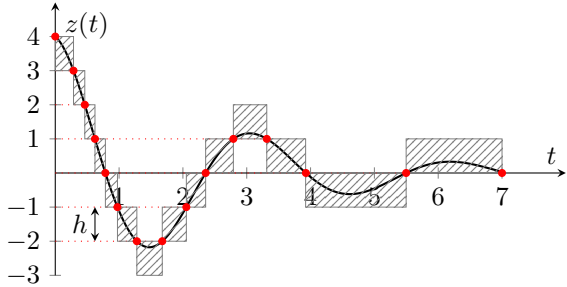
\begin{figure}
	\begin{center}
		\begin{tikzpicture}
			[
			declare function={
				func1(\x)= 	4*exp(-\x/2.5)*cos(deg(2*\x));}
			]
			\begin{axis}[
				width=8.3cm, height=5.2cm, axis x line=middle, axis y line=middle,
				ymin=-3, ymax=5, ytick={-3,...,4}, ylabel=$z(t)$,
				xmin=0, xmax=8, xtick={0,...,7}, xlabel=$t$,
				domain=-1:8,samples=301, 
				]
				\addplot [black,thick,domain=0:7] {func1(x)};
				\addplot [dotted,red] coordinates {(0,3) (0.2859,3)};
				\addplot [dotted,red] coordinates {(0,2) (0.4626,2)};
				\addplot [dotted,red] coordinates {(0,1) (3.314,1)};
				\addplot [dotted,red] coordinates {(0,0) (5.4977,0)};
				\addplot [dotted,red] coordinates {(0,-1) (2.0538,-1)};
				\addplot [dotted,red] coordinates {(0,-2) (1.6765,-2)};
				\draw[Rect] (0,600) rectangle (28.59,700);
				\draw[Rect] (28.59,500) rectangle (46.26,600);
				\draw[Rect] (46.26,400) rectangle (62.22,500);
				\draw[Rect] (62.22,300) rectangle (78.54,400);
				\draw[Rect] (78.54,200) rectangle (97.45,300);
				\draw[Rect] (97.45,100) rectangle (127.83,200);
				\draw[Rect] (127.83,0) rectangle (167.65,100);
				\draw[Rect] (167.65,100) rectangle (205.38,200);
				\draw[Rect] (205.38,200) rectangle (235.62,300);
				\draw[Rect] (235.62,300) rectangle (279.04,400);
				\draw[Rect] (279.04,400) rectangle (331.40,500);
				\draw[Rect] (331.40,300) rectangle (392.27,400);
				\draw[Rect] (392.27,200) rectangle (549.77,300);
				\draw[Rect] (549.77,300) rectangle (700,400);
				\draw [stealth-stealth](62.22,100) -- (62.22,200);
				\node[text width=3cm] at (205,150) 
				{$h$};
				\addplot [only marks,mark=*,red,mark size=1.3] coordinates {(0,4)(0.2859,3)(0.4626,2)(0.6222,1)(2.7904,1)(3.3140,1)(0.7854,0)(2.3562,0)(0.9745,-1)(3.927,0)(5.4977,0)(2.0538,-1)(1.2783,-2)(1.6765,-2)(7,0)};
			\end{axis}
		\end{tikzpicture}
	\end{center}
	\vspace{-0.4cm}
	\caption{Lebesgue sampling of a signal $z(t)$ with threshold $h = 1$. Red dots indicate the sampling instants and thresholds being crossed, and dashed gray rectangles show the regions where $z(t)$ is known to be located.}
	\label{fig1}
\end{figure}
\hspace{-0.13cm}with $m_l = z(t_l)/h$. Without loss of generality, we assume that $t_1=0$. We pursue a non-parametric estimate of $g$ using the continuous-time input $\{u(t)\}_{t\in [0,t_{N_\textnormal{L}}]}$ and the Lebesgue-sampled output $\{y_\textnormal{L}(t_l)\}_{l=1}^{N_\textnormal{L}}$. 

\subsection{Practical framework for Lebesgue-sampled systems}
Incremental encoders operate on this kind of sampling principle \citep{merry2013optimal}. In practice, a light source emits a beam directed towards a slotted disk or strip, and the output of two light detectors are recorded. These two signals allow the encoder to detect the direction of the rotation. The quantity $h$ represents the uncertainty in the measurements of the incremental encoder, which is inversely proportional to its resolution. In low-resolution incremental encoders, the quantization effect produced by $h$, in conjunction with the non-equidistant nature of the sampling mechanism, can impact the performance of, e.g., iterative learning control~\citep{strijbosch2022iterative}.

Behind this sampling procedure there typically is an amplitude detection mechanism that operates at a fast sampling rate. With this context in mind, we define $\Delta>0$ as the sampling period of this amplitude detection mechanism. The following assumption is set in place.
\begin{assum}
	\label{assumption12}
	For every time instant $t=i\Delta$, $i=0,1,\dots,\lfloor t_{N_\textnormal{L}}/\Delta\rfloor +1$, the lower and upper threshold levels associated with the unsampled output $z(t)$ are known. The lower bound at each time-instant $t=i\Delta$ is denoted as $\eta_i$, and it can be deduced unambiguously from $\{y_{\textnormal{L}}(t_l)\}_{l=1}^{N_\textnormal{L}}$. 
\end{assum}

Thus, a \textit{set-valued} signal that represents a fast-sampled version of the gray rectangles in Fig. \ref{fig1} can be defined as
\begin{equation}
	\label{qh}
	y(i\Delta) = \mathcal{Q}_h\{z(i\Delta)\}:=[\eta_i,\eta_i+h).
\end{equation}
In the proposed setup, the practitioner has access to $y(i\Delta)$ as output information. To simplify our exposition (and with some abuse of notation), we denote $\{y(i\Delta)\}_{i=0}^{N}$ and $\{z(i\Delta)\}_{i=0}^{N}$ as $\mathbf{y}_{0:N}$ and $\mathbf{z}_{0:N}$, respectively, with $N:=\lfloor t_{N_\textnormal{L}}/\Delta\rfloor +1$. Note that all vectors and matrices are written in bold throughout the paper.

Note that we do not assume that each time-stamp $t_l$ is a multiple of $\Delta$. Although such assumption is commonly used in intermittent sampling setups \citep{strijbosch2022iterative}, we do not require it in this paper.

\begin{assum}
	\label{assumption13}
	At every time instant $t=i\Delta$, the disturbance noise affecting the output $z(i\Delta)$ is an additive discrete-time independent and identically distributed (i.i.d.) Gaussian white noise of variance $\sigma^2$.
\end{assum}

The noise variance is not known beforehand and is typically estimated from the data or selected \textit{a priori} according to expert knowledge. After taking into consideration Assumptions \ref{assumption12} and \ref{assumption13}, the problem of interest is: Considering the continuous-time input $\{u(t)\}_{t\in [0,t_{N_\textnormal{L}}]}$ and the set-valued data $\{y(i\Delta)\}_{i=0}^{N}$, obtain a non-parametric estimate of the underlying continuous-time impulse response $g$.
\begin{remark}
	\label{remark21}
	In many cases, the input $u$ is generated by a zero-order-hold device of sampling period $\Delta_u$. When $\Delta_u$ is a multiple of $\Delta$, we may consider the sampled input signal $\{u(i\Delta)\}_{i=0}^{N}$ instead of a fully continuous-time description. Both viewpoints are equivalent if the intersample behavior of the sampled input is known and correctly incorporated in the construction of the algorithms.
\end{remark}
\begin{remark}
	We will only consider the output data that are produced by the input excitation starting from $t=0$. We discard the first output measurement $y(0)$ due to causality.
\end{remark}

\section{Non-parametric estimation using Lebesgue-sampled data}
\label{sec:kernel}
This paper solves the identification problem by incorporating the Lebesgue-sampling strategy into the framework of continuous-time kernel-based methods. This section addresses the main challenges regarding how to adequately exploit the output uncertainty set information instead of point data, and how to tune the kernel hyperparameters. 

\subsection{MAP estimator for Lebesgue-sampled systems}
\label{section3A}
We begin by addressing the Lebesgue-sampled system identification problem from a Bayesian standpoint. In this regard, we assume that $g$ can be modeled by a zero-mean Gaussian process with prior distribution $\textnormal{p}(g)$ and covariance $\mathbb{E}\{g(t)g(s)\} = \gamma^{-1}k(t,s)$ with $\gamma>0$, and we are interested in computing the MAP-inspired estimator
\begin{equation}
	\label{mapestimator}
	\hat{g}_{\textnormal{MAP}}(t) = \underset{g}{\arg \max} \big(\ell(g) + \log \textnormal{p}(g)\big),
\end{equation}
where $\ell(\cdot)$ denotes the log-likelihood function $\ell(g) = \log \textnormal{p}(\mathbf{y}_{1:N}|g)$, with $\textnormal{p}$ denoting the probability density function. The prior $\textnormal{p}(g)$ must be described in some manner so that \eqref{mapestimator} is tractable. Although the prior density cannot be formally defined in infinite-dimensional function spaces, there exists a formal connection \citep{aravkin2014connection} between MAP estimation and kernel methods by setting the prior density being proportional to $\exp(-\gamma \|g\|^2_{\mathcal{G}})$, where $\mathcal{G}$ is a reproducing kernel Hilbert space (RKHS) that has $k$ as its kernel. Thus, the first result we present concerns finding transparent expressions for
\begin{equation}
\label{kernelbased}
\hat{g} = \underset{g\in \mathcal{G}}{\arg \min} \big(\hspace{-0.04cm}-\log \textnormal{p}(\mathbf{y}_{1:N}|g) + \gamma \|g\|_{\mathcal{G}}^2\big),
\end{equation} 
for a \textit{fixed} kernel $k$, when $\mathbf{y}_{1\hspace{-0.02cm}:\hspace{-0.02cm}N}$ is Lebesgue-sampled data. The next\hspace{0.09cm}theorem\hspace{0.09cm}constitutes\hspace{0.08cm}Contribution\hspace{0.06cm}C1\hspace{0.06cm}of\hspace{0.09cm}this\hspace{0.09cm}work.

\begin{thm}
	\label{thm1}
Under Assumptions \ref{assumption12} and \ref{assumption13}, the kernel-based estimator \eqref{kernelbased} is given by
\begin{equation}
	\label{finiterepresentation}
	\hat{g}(t) = \textstyle\sum_{i=1}^{N} c_i \hat{g}_i(t),
\end{equation}
where 
\begin{equation}
	\label{indeed}
	\hat{g}_i(t) = \int_0^\infty u(i\Delta-\tau)k(t,\tau)\textnormal{d}\tau,
\end{equation}
and the vector of coefficients $\hat{\mathbf{c}}:=[c_1, c_2,\dots, c_{N}]^\top$ is obtained by solving the optimization problem
	\begin{equation}
	\label{computec}
	\hspace{-0.15cm}\hat{\mathbf{c}} \hspace{-0.07cm}= \hspace{-0.05cm} \underset{\mathbf{c}\in \mathbb{R}^{N}}{\arg \min} \hspace{-0.08cm} \left(\hspace{-0.12cm}-\hspace{-0.06cm} \sum_{i=1}^{N}\hspace{-0.04cm} \log \hspace{-0.09cm}\left[\hspace{-0.05cm}\int_{\eta_i}^{\eta_i\hspace{-0.03cm}+\hspace{-0.02cm}h}\hspace{-0.31cm} e^{\frac{-1}{2\sigma^2}\hspace{-0.05cm}\big(\hspace{-0.03cm}z_i\hspace{-0.03cm}-\hspace{-0.03cm}\mathbf{K}_i^{\hspace{-0.03cm}\top} \hspace{-0.06cm}\mathbf{c}\big)^{\hspace{-0.02cm}2}}\hspace{-0.09cm}\textnormal{d}z_i \hspace{-0.02cm}\right]\hspace{-0.13cm}+\hspace{-0.08cm} \gamma \mathbf{c}^{\hspace{-0.03cm}\top}\hspace{-0.03cm} \mathbf{Kc}\hspace{-0.05cm} \right)\hspace{-0.07cm},
\end{equation}
with\hspace{0.09cm}$\mathbf{K}_i$\hspace{0.09cm}being\hspace{0.09cm}the\hspace{0.09cm}$i$th\hspace{0.09cm}column\hspace{0.09cm}of\hspace{0.09cm}the\hspace{0.09cm}matrix\hspace{0.09cm}$\mathbf{K}$\hspace{0.09cm}with entries
\begin{equation}
	\label{kernelmatrix}
	\mathbf{K}_{ij} = \int_0^\infty \hspace{-0.2cm}\int_0^\infty u(i\Delta-\xi) u(j\Delta-\tau)k(\xi,\tau)\textnormal{d}\tau \textnormal{d}\xi.
\end{equation}
\end{thm}

\begin{pf}
We first obtain an expression for $\textnormal{p}(\mathbf{y}_{1:N}|g)$. The probability density function of the output prior to sampling $\mathbf{z}_{1\hspace{-0.02cm}:\hspace{-0.02cm}N}\hspace{-0.09cm}=\hspace{-0.1cm}[z(\Delta),\hspace{-0.02cm}\dots\hspace{-0.02cm},\hspace{-0.02cm}z(N\Delta)]^{\top}\hspace{-0.04cm}$ (conditioned to $g$) is
\begin{equation}
	\textnormal{p}\left(\mathbf{z}_{1:N}|g\right) = \frac{1}{(2\pi \sigma^2)^{\frac{N}{2}}} \prod_{i=1}^{N} e^{-\frac{1}{2\sigma^2}\big[z(i\Delta)-(g*u)(i\Delta)\big]^2}, \notag
\end{equation}
where we have used the fact that the additive noise is Gaussian and i.i.d. by Assumption \ref{assumption13}. Therefore, from \eqref{qh}, the probability mass function of $\mathbf{y}_{1:N}$ is
\begin{align}
	\hspace{-0.2cm}\textnormal{p}\hspace{-0.05cm}\left(\hspace{-0.02cm}\mathbf{y}_{1\hspace{-0.015cm}:\hspace{-0.015cm}N}\hspace{-0.02cm}|g\hspace{-0.01cm}\right) & \hspace{-0.08cm}=\hspace{-0.08cm}  \mathbb{P} \hspace{-0.01cm}\Big(\hspace{-0.03cm}z(\hspace{-0.02cm}\Delta\hspace{-0.02cm})\hspace{-0.07cm}\in\hspace{-0.07cm}[\eta_1\hspace{-0.02cm},\hspace{-0.02cm}\eta_1\hspace{-0.06cm}+\hspace{-0.05cm}h),\hspace{-0.02cm}\dots\hspace{-0.02cm},\hspace{-0.02cm}z(\hspace{-0.02cm} N \hspace{-0.02cm}\Delta\hspace{-0.02cm} )\hspace{-0.07cm}\in\hspace{-0.07cm} [\eta_{N}\hspace{-0.02cm},\hspace{-0.02cm} \eta_{N}\hspace{-0.06cm}+\hspace{-0.05cm} h)|g\hspace{0.02cm}\Big) \notag \\
	\label{eq5}
	&\hspace{-0.08cm}=\hspace{-0.09cm} \frac{1}{(2\pi \sigma^2\hspace{-0.02cm})^{\hspace{-0.04cm}\frac{N}{2}}}\hspace{-0.03cm} \prod_{i=1}^{N} \hspace{-0.03cm}\int_{\eta_i}^{\eta_{i}\hspace{-0.03cm}+\hspace{-0.03cm}h} \hspace{-0.1cm}e^{-\frac{1}{2\sigma^2}\big[z_i-(g*u)(i\Delta)\big]^2}\textnormal{d}z_i.
\end{align}
Therefore, computing \eqref{kernelbased} is equivalent to solving
\begin{equation}
	\label{regularization2}
	\min_{g\in \mathcal{G}} \hspace{-0.08cm}\left(\hspace{-0.11cm}-\hspace{-0.06cm}\sum_{i=1}^{N}\hspace{-0.04cm} \log \hspace{-0.07cm}\left[\hspace{-0.05cm}\int_{\eta_i}^{\eta_i\hspace{-0.03cm}+\hspace{-0.02cm}h}\hspace{-0.24cm} e^{\frac{-1}{2\sigma^2}\big[z_i-(g*u)(i\Delta)\big]^2}\hspace{-0.05cm}\textnormal{d}z_i \right] \hspace{-0.1cm}+\hspace{-0.07cm} \gamma \hspace{-0.03cm}\|g\|_\mathcal{G}^2 \hspace{-0.05cm}\right)\hspace{-0.1cm}.
\end{equation}
By the representer theorem \citep{scholkopf2001generalized}, any $g\in \mathcal{G}$ that minimizes the cost in \eqref{regularization2} admits a representation of the form \eqref{finiterepresentation} with representers $\hat{g}_i$ in \eqref{indeed}. Note that any minimizer $g$ satisfies $(g*u)(i\Delta) = \mathbf{K}_i^\top \mathbf{c}$, with $\mathbf{K}_i$ being the $i$th column of the kernel matrix described by \eqref{kernelmatrix}, and also $\|g\|^2_\mathcal{G} = \mathbf{c}^\top \mathbf{K}\mathbf{c}$. Thus, replacing the finite-dimensional representation \eqref{finiterepresentation} in \eqref{regularization2} reduces the optimization problem to finding $\mathbf{c}$ that minimizes \eqref{computec}. \hspace{2.5cm} \hfill \qed 
\end{pf}
\begin{remark}
	Although $\hat{g}$ in \eqref{finiterepresentation} is not strictly speaking a MAP estimator of the impulse response $g$, it can be shown that $(\hat{g}*u)$ provides a MAP estimator of the output prior to Lebesgue sampling at the instants $t=i\Delta$, $i=1,\dots,N$. The proof of this result, which complements the one in \cite{aravkin2014connection} for Riemann-sampled data, is out of the scope of this paper and will be published elsewhere.
\end{remark} 

The main contribution of Theorem \ref{thm1} is recasting the infinite-dimensional problem for Lebesgue-sampled system identification in \eqref{kernelbased} as a finite dimensional one in \eqref{computec}. Note that $\hat{\mathbf{c}}$ does not have an explicit form as the Riemann-sampling counterpart \citep{pillonetto2022regularized}, where it is computed directly as the solution of a regularized least-squares problem. The following subsection is focused on how to compute $\hat{\mathbf{c}}$ for a fixed kernel $k$ and hyperparameters $\gamma$ and $\sigma^2$. Later, we present an algorithm for computing the kernel hyperparameters using Empirical Bayes.

\subsection{Optimal weights with MAP-EM}
To compute \eqref{computec}, we propose an iterative procedure based on the MAP-EM algorithm. This procedure, which ensures the computation of a local maximum of the cost in \eqref{computec} under general conditions as a extension of the standard EM \citep{wu1983convergence}, constitutes Contribution C2 of this paper. The approach exploits the fact that \eqref{computec} is related to the MAP estimator of an FIR model, which is used for applying the EM algorithm tailored for MAP estimation. The following lemma makes this connection evident.
\begin{lem}
	\label{lemma32}
	Consider the following linear regression model with a set-valued output:
	\begin{subequations}
		\label{modelforc}
		\begin{align}
			\label{modelforc1}
			z(i\Delta) &= \mathbf{K}_i^\top \mathbf{c} + e(i\Delta),  \\
			\label{modelforc2}
			y(i\Delta) &= \mathcal{Q}_h\{z(i\Delta)\}, 
		\end{align}
	\end{subequations}
	where $\{e(i\Delta)\}_{i=1}^N$ is an i.i.d. Gaussian white noise with variance $\sigma^2$, and $\mathbf{K}_i, i=1,2,\dots,N$, is assumed known. Assume that $\mathbf{c}$ in \eqref{modelforc1} has a Gaussian prior distribution, with zero mean and covariance $\mathbf{K}^{-1}/(2\gamma)$. Then, the MAP estimator for $\mathbf{c}$ is given by $\hat{\mathbf{c}}$ in \eqref{computec}.
\end{lem}
\begin{pf}
	See Appendix \ref{prooflemma32}. \hspace{4.2cm} \hfill \qed 
\end{pf}

By Lemma \ref{lemma32} we can view the computation of the weights $\hat{\mathbf{c}}$ in a MAP-EM framework if we set the unquantized output $z(i\Delta)$ in \eqref{modelforc1} as our hidden variable. In other words, we can optimize the\textit{ a posteriori} density for $\mathbf{c}$ (which is exactly the objective function in \eqref{computec}) by computing the conditional expectation (i.e., the E-step) of the log complete-data posterior density given the measurements $\mathbf{y}_{1:N}$ and the current estimate of $\hat{\mathbf{c}}$, and later performing a maximization step (i.e., the M-step). These two steps are outlined in Algorithm~\ref{algorithm1}. Details pertaining the EM algorithm and its derivation can be found in, e.g., \citep{dempster1977maximum}.
\begin{algorithm}
	\caption{MAP-EM for computing $\hat{\mathbf{c}}$ in \eqref{computec}}
	\begin{algorithmic}[1]
		\State Input: initial estimate $\hat{\mathbf{c}}^{(1)}$
		\For{$j = 1, 2, \dots$}
		\State \textbf{E-step}: Compute the expectation
		\begin{equation}
			\label{qfunction}
			Q(\mathbf{c},\hspace{-0.02cm}\hat{\mathbf{c}}^{(j)}\hspace{-0.02cm}) \hspace{-0.05cm}= \hspace{-0.04cm}\mathbb{E}\hspace{-0.03cm}\big\{ \hspace{-0.04cm} \log \textnormal{p}(\mathbf{z}_{1:N}\hspace{-0.01cm},\hspace{-0.02cm}\mathbf{y}_{1:N}|\mathbf{c})|\mathbf{y}_{1:N}\hspace{-0.01cm},\hspace{-0.01cm}\hat{\mathbf{c}}^{(j)}\hspace{-0.02cm}\big\}\hspace{-0.06cm}
		\end{equation} 
		\State \textbf{M-step}: Solve the optimization problem
		\begin{equation}
			\label{mstep}
			\hat{\mathbf{c}}^{(j+1)} = \underset{\mathbf{c}\in \mathbb{R}^N}{\arg \max} \left( Q(\mathbf{c},\hat{\mathbf{c}}^{(j)}) -\hspace{-0.07cm}\gamma\mathbf{c}^{\hspace{-0.06cm}\top}\hspace{-0.03cm} \mathbf{Kc} \right)
		\end{equation}
			\vspace{-0.3cm}
		\EndFor
	\end{algorithmic}
	\label{algorithm1}
\end{algorithm}

The E-step is obtained from a result from quantized FIR estimation. Afterwards, we present the M-step in Theorem~\ref{theorem31}, which contains the iterations for computing~$\hat{\mathbf{c}}$.
\begin{lem}{\cite[Lemma 5]{godoy2011identification}}.
	\label{lemma33}
	Consider the discrete-time model \eqref{modelforc}. The $Q$ function in \eqref{qfunction} satisfies
	\begin{equation}
		Q\hspace{-0.02cm}(\mathbf{c},\hspace{-0.03cm}\hat{\mathbf{c}}^{(\hspace{-0.01cm}j\hspace{-0.01cm})}\hspace{-0.04cm}) \hspace{-0.09cm}= \hspace{-0.09cm} \hspace{-0.02cm}\frac{-1}{2\sigma^2} \hspace{-0.07cm} \sum_{i=1}^{N}\hspace{-0.06cm} \int_{\eta_i}^{\eta_i\hspace{-0.02cm}+\hspace{-0.02cm}h} \hspace{-0.27cm}(z_i\hspace{-0.05cm}-\hspace{-0.05cm}\mathbf{K}_i^\top \hspace{-0.03cm}\mathbf{c})^{\hspace{-0.02cm}2} \textnormal{p}(z_i|y(\hspace{-0.01cm}i\Delta \hspace{-0.01cm}),\hat{\mathbf{c}}^{(j)}\hspace{-0.03cm})\textnormal{d}z_i+C, \notag
	\end{equation}
	where $C$ is a constant.
\end{lem}
\begin{pf}
	See \cite{godoy2011identification}. \hspace{3.2cm} \hfill \qed 
\end{pf}
\begin{thm}
	\label{theorem31}
	The M-step in \eqref{mstep} is equivalent to
	\begin{equation}
		\label{ck1}
		\hat{\mathbf{c}}^{(j+1)} = (\mathbf{K}+ \tilde{\gamma} \mathbf{I})^{-1} \tilde{\mathbf{z}}^{(j)},
	\end{equation}
	where $\tilde{\gamma} = 2\sigma^2\gamma$, and with the $i$th entry of $\tilde{\mathbf{z}}^{(j)}$ being
	\begin{equation}
		\label{tildezi}
		\hspace{-0.2cm}\tilde{z}_i^{(j)} \hspace{-0.14cm} = \hspace{-0.07cm}\mathbf{K}_i^{\hspace{-0.04cm}\top} \hspace{-0.04cm}\hat{\mathbf{c}}^{\hspace{-0.02cm}(j)} \hspace{-0.1cm}+ \frac{\hspace{-0.18cm}\sqrt{\hspace{-0.07cm}\frac{2}{\pi}}\sigma \hspace{-0.08cm}\left(\hspace{-0.07cm}\exp\hspace{-0.06cm}\big\{\hspace{-0.15cm}-\hspace{-0.09cm}(\hspace{-0.02cm}b_i^{\hspace{-0.03cm}(j)}\hspace{-0.03cm})^2 \hspace{-0.04cm}\big\} \hspace{-0.11cm}-\hspace{-0.07cm} \exp\hspace{-0.06cm}\big\{\hspace{-0.15cm}-\hspace{-0.09cm}(\hspace{-0.02cm}b_i^{\hspace{-0.03cm}(j)}\hspace{-0.15cm}+\hspace{-0.1cm}\frac{h}{\sqrt{\hspace{-0.02cm}2} \sigma}\hspace{-0.03cm})^2 \hspace{-0.04cm}\big\} \hspace{-0.07cm} \right)}{\textnormal{erf}\big[b_i^{(j)}\hspace{-0.07cm}+\hspace{-0.07cm}\frac{h}{\sqrt{2} \sigma}\big]-\textnormal{erf}\big[b_i^{(j)}\big]}\hspace{-0.02cm},
	\end{equation}
	where $b_i^{(j)} := (\eta_i-\mathbf{K}_i^\top \hat{\mathbf{c}}^{(j)})/(\sqrt{2} \sigma)$, and the error function is defined by $\textnormal{erf}[x]:=(2/\sqrt{\pi}) \int_{0}^{x} \textnormal{exp}(-t^2)\textnormal{d}t$.
\end{thm}
\begin{pf}
	See Appendix \ref{prooftheorem31}. \hspace{4.2cm} \hfill \qed 
\end{pf}
\subsection{Kernel hyper-parameter optimization}
\label{subD}
The optimal weighting $\hat{\mathbf{c}}$ depends on the kernel that is selected (together with the regularization constant $\gamma$), and of the noise covariance $\sigma^2$. These unknown quantities are typically encompassed in a hyperparameter vector $\bm{\rho}$ that must be estimated using the data at hand. We pursue an Empirical Bayes (EB) approach to estimate these unknown parameters. The EB algorithm is highly efficient, as is evidenced in its broad application \citep{pillonetto2014kernel,scandella2022kernel}.

There are many well-known kernels that are used for continuous-time impulse response estimation. For example, the stable-spline one of order $q$ is defined as
\begin{equation}
	k(t,\tau) = s_q(e^{-\beta t},e^{-\beta \tau}), \quad q\in\mathbb{N},\notag  
\end{equation}
where $\beta$ is a positive hyperparameter, and $s_q$ is the regular spline kernel of order $q$. For $q=1$ we have $s_{1}(e^{-\beta t},e^{-\beta \tau})= e^{-\beta \max(t,\tau)}$. The EB approach involves estimating the vector $\bm{\rho}=[\gamma, \beta, \sigma^2]^\top$ by solving the optimization problem
\begin{equation}
	\label{eb}
	\hat{\bm{\rho}}_{\textnormal{EB}} = \underset{\bm{\rho}\in \bm{\Gamma}}{\arg \max} \hspace{0.1cm}\textnormal{p}(\mathbf{y}_{1:N}|\bm{\rho}),  
\end{equation}
where $\bm{\Gamma}$ denotes the admissible space of hyperparameters (i.e., $\gamma,\beta,\sigma^2>0$ in the stable-spline case). To describe such optimization problem more explicitly, we first compute the probability density function of the output prior to Lebesgue sampling, $\textnormal{p}(\mathbf{z}_{1:N}|\bm{\rho})$. This expression can be obtained directly by exploiting the fact that the additive noise is assumed Gaussian and independent of $g$ (which is also assumed Gaussian), which leads to
\begin{equation}
	\label{zrho}
	\mathbf{z}_{1:N}|\bm{\rho} \sim \mathcal{N}(\mathbf{0},\mathbf{K}_\beta/\gamma+\sigma^2 \mathbf{I}),
\end{equation}
where we have made explicit the dependence of the kernel matrix $\mathbf{K}$ on the hyperparameter $\beta$. Therefore, the EB estimator for $\bm{\rho}$ is given by
\begin{align}
	\hat{\bm{\rho}}_{\textnormal{EB}} &= \underset{\bm{\rho} \in \bm{\Gamma}}{\arg \max} \frac{1}{\sqrt{\det(2\pi[\mathbf{K}_\beta/\gamma+\sigma^2 \mathbf{I}])}} \notag \\
	\label{kernelopt}
	&\times \int_{\mathbf{z}\in \mathbf{y}_{1:N}} \hspace{-0.32cm}\exp\left\{\hspace{-0.07cm}-\hspace{-0.03cm}\frac{1}{2}\mathbf{z}^\top (\mathbf{K}_\beta/\gamma \hspace{-0.07cm}+ \hspace{-0.06cm}\sigma^2 \mathbf{I})^{-1}\mathbf{z} \hspace{-0.03cm}\right\} \hspace{-0.04cm}\textnormal{d}\mathbf{z},
\end{align}
where the elements of $\mathbf{y}_{1:N}$ are defined in \eqref{qh}. This optimization problem involves an $N$-dimensional integral, which is hard to compute in general (see, e.g., \cite{chen2012impulse,bottegal2017new}). The intractability is here addressed by optimizing \eqref{kernelopt} with EM, similarly as in the previous subsection. For brevity, we derive the EM iterations jointly (both E and M steps) in Theorem \ref{theorem32}.
\begin{thm}
	\label{theorem32}
	The following iterative procedure is guaranteed to converge with probability 1 to a (local or global) maximum for the cost in \eqref{kernelopt}:
	\begin{equation}
		\label{emiterationsthm32}
		\hat{\bm{\rho}}^{(j+1)} \hspace{-0.07cm}=\hspace{-0.07cm} \underset{\bm{\rho} \in \bm{\Gamma}}{\arg \min} \left(\log \det(\mathbf{S}_{\bm{\rho}})+\textnormal{tr}\{\mathbf{S}_{\bm{\rho}}^{-1} \bar{\mathbf{Q}}^{(j)}\hspace{-0.02cm}\}\right),
	\end{equation}
	where $\mathbf{S}_{\bm{\rho}}:= \mathbf{K}_\beta/\gamma + \sigma^2 \mathbf{I}$, and $\bar{\mathbf{Q}}^{(j)}$ is the second moment of $\mathbf{z}_{1:N}$ given the data and the $j$th iteration of $\hat{\bm{\rho}}$, i.e.,
	\begin{equation}
		\label{secondmoment}
		\bar{\mathbf{Q}}^{(j)} = \mathbb{E}\{\mathbf{z}_{1:N}\mathbf{z}_{1:N}^\top |\mathbf{y}_{1:N},\hspace{-0.02cm}\hat{\bm{\rho}}^{(j)}\hspace{-0.04cm}\}.  
	\end{equation}
\end{thm}
\begin{pf}
	See Appendix \ref{prooftheorem32}. \hspace{4.2cm} \hfill \qed 
\end{pf}
	The iterations in Theorem \ref{theorem32} reduce the $N$-dimensional integral optimization problem into a form similar to the Empirical Bayes optimization problem \citep{pillonetto2014kernel} for unquantized output data, at the expense of having to compute the second moment of $\mathbf{z}_{1:N}$ conditioned on the set-valued data. The $\bar{\mathbf{Q}}^{(j)}$ matrix cannot be computed in closed-form in general. In this paper, we extract samples of a multivariate truncated Gaussian distribution using the algorithm in \cite{botev2017normal} and we approximate the expectation in \eqref{secondmoment} via Monte Carlo integration.

\subsection{The complete algorithm}
To conclude this section, the full algorithm for non-parametric identification of Lebesgue-sampled continuous-time systems is described in Algorithm \ref{algorithm2}.
\begin{algorithm}
	\caption{Kernel-based non-parametric identification for Lebesgue-sampled continuous-time systems}
	\begin{algorithmic}[1]
		\State Input: $\mathbf{u}_{0:N-1},\mathbf{y}_{1:N}$, initial hyperparameter estimate $\hat{\bm{\rho}}^{(1)}=[\hat{\gamma}^{(1)}, \hat{\beta}^{(1)},{\hat{\sigma}^2{}}^{(1)}]^\top$, initial weighting estimate $\hat{\mathbf{c}}^{(1)}$, maximum number of MAP-EM iterations $M$
		\For{$j = 1, 2, \dots,M$}
		\State Compute $\bar{\mathbf{Q}}^{(j)}$ from \eqref{secondmoment} using \cite{botev2017normal} 
		\State Obtain $\hat{\bm{\rho}}^{(j+1)} \hspace{-0.1cm}=\hspace{-0.1cm} [\hat{\gamma}^{(j+1)}, \hat{\beta}^{(j+1)},{\hat{\sigma}^2{}}^{(j+1)}]^\top$ from \eqref{emiterationsthm32}
		\EndFor
		\For{$j = 1, 2, \dots,M$}
		\State Obtain $\hat{\mathbf{c}}^{(j+1)}$ from \eqref{ck1} with $\mathbf{K}$, $\tilde{\gamma}$ and $\tilde{\mathbf{z}}^{(j)}$ com-\- \hspace*{0.55cm}puted using $\hat{\bm{\rho}}^{(M+1)}$
		\EndFor
		\State Output: $\hat{g}$ obtained from \eqref{finiterepresentation}, with $\hat{g}_i$ computed from \eqref{indeed} using $\hat{\beta}^{(M+1)}$ as the hyperparameter for $k$.		
	\end{algorithmic}
	\label{algorithm2}
\end{algorithm}
\section{Case study}
\label{sec:simulations}
In this section, the novel non-parametric estimator is tested on a practically-relevant example. We consider a mass-spring-damper system with transfer function given~by
\begin{equation}
	G(s) = \frac{1}{\textnormal{m}s^2+\textnormal{d}s+\textnormal{k}}, \notag
\end{equation}
with mass $\textnormal{m}\hspace{-0.07cm}=\hspace{-0.07cm}0.05$[Kg], damping coefficient $\textnormal{d}\hspace{-0.07cm}=\hspace{-0.07cm}0.2$[Ns/m], and spring constant $\textnormal{k}\hspace{-0.07cm}=\hspace{-0.07cm}1$[N/m]. The output is sensed with period $\Delta\hspace{-0.07cm}=\hspace{-0.07cm}0.1$[s], and $h\hspace{-0.07cm}=\hspace{-0.07cm}1$[m]. The input is a Gaussian white noise sequence passed through a zero-order hold device with period $\Delta_u=3$[s]. One hundred Monte Carlo runs are performed with a varying input and an additive Gaussian white noise prior to the Lebesgue-sampling with standard deviation $0.05$[m]. Each run has a total time duration of $30$[s] (i.e., $300$ data points are sensed prior to Lebesgue-sampling), and on average $N_{\textnormal{L}}=69$ output samples are obtained after sampling per run.

Three estimators are tested: the standard kernel-based continuous-time non-parametric estimator \citep{scandella2022kernel} using the midpoint estimate $z(i\Delta)\approx \eta_i+h/2$ as output data ($\hat{g}_{\textnormal{rie}}$), this same estimator but using the noisy output $z(i\Delta)$ prior to Lebesgue sampling as output data ($\hat{g}_{\textnormal{or}}$), and the proposed approach (Algorithm \ref{algorithm2} of this paper, $\hat{g}_{\textnormal{leb}}$). Note that the oracle estimator $\hat{g}_{\textnormal{or}}$ cannot be implemented in practice, since we do not have direct knowledge of the system output before the event-sampler. We measure each estimator's performance with the metric
\begin{equation}
	\textnormal{fit} = 100 \big(1- \|\hat{\mathbf{x}}^j-\mathbf{x}\|_2/\|\mathbf{x}-\bar{x}\mathbf{1}\|_2 \big), \notag
\end{equation}
where $\mathbf{x}$ is the noiseless output sequence (prior to Lebesgue-sampling), $\hat{\mathbf{x}}^j$ is the simulated output using the $j$th impulse response estimate, and $\bar{x}$ is the mean value of $\mathbf{x}$. The proposed estimator uses the 1st order stable-spline kernel with a maximum number of MAP-EM iterations $M=40$, and $1000$ samples of a multivariate truncated Gaussian distribution are obtained to compute $\bar{\mathbf{Q}}^{\hspace{-0.02cm}(\hspace{-0.01cm}j\hspace{-0.01cm})}$\hspace{0.05cm}in\hspace{0.05cm}\eqref{secondmoment}.

\begin{figure}
	\centering{
		\includegraphics[width=0.45\textwidth]{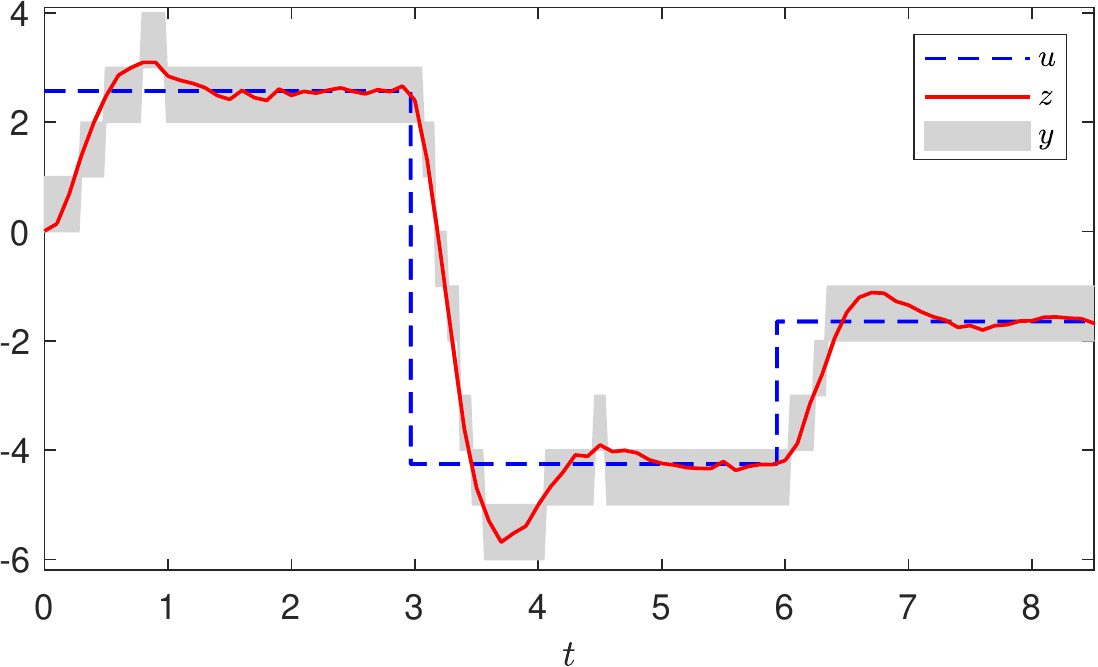}
		\vspace{-0.23cm}
		\caption{Input and output signals corresponding to $8$[s] of one Monte Carlo run, with $h=1$. The proposed estimator $\hat{g}_{\textnormal{leb}}$ only uses the fact that the unsampled output is located in the grey band $y$.}
		\label{fig2}}
\end{figure}

\begin{figure}
	\centering{
		\includegraphics[width=0.37\textwidth]{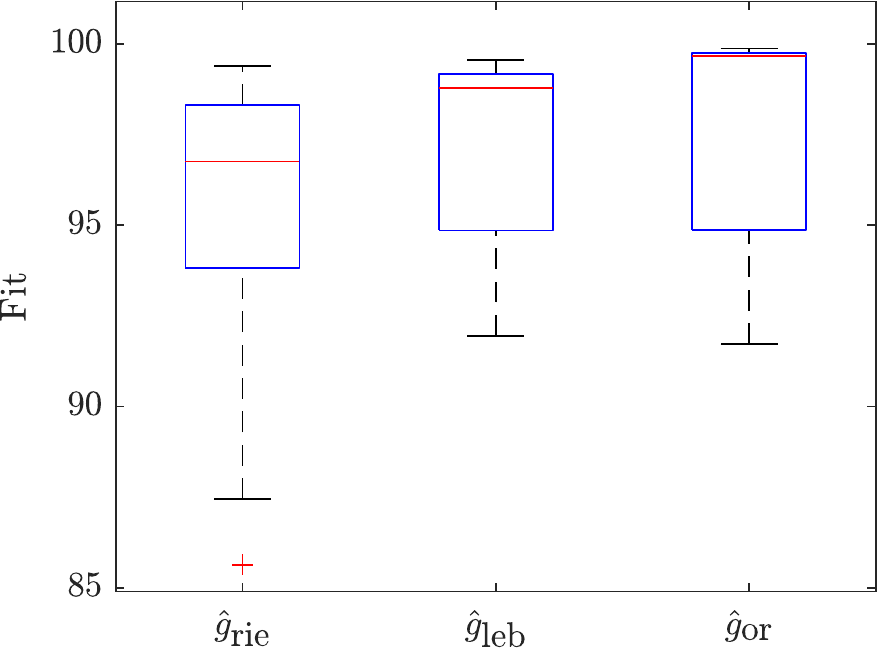}
		\vspace{-0.2cm}
		\caption{Boxplots of the fit metric for the case study. The proposed estimator $\hat{g}_{\textnormal{leb}}$ outperforms the midpoint method $\hat{g}_{\textnormal{rie}}$, and is close in performance to the oracle estimator $\hat{g}_{\textnormal{or}}$, despite using fewer effective samples.}
		\label{fig3}}
\end{figure}

In Figure \ref{fig2}, a typical data set is shown. Note that the task of the proposed estimator is particularly challenging, since the overshoot of the output signal $z$ is rarely captured in the $y$ signal band due to the coarse grid produced by the threshold level. Figure \ref{fig3} shows the boxplots of the fit metric for each estimator. The proposed approach achieves on average a better fit than the estimator that only uses the midpoint values $\eta_i+h/2$ as output. The $\hat{g}_{\textnormal{leb}}$ estimator is only slightly outperformed by the oracle, despite having a low resolution for the output measurement mechanism and a $77\%$ reduction in output data samples on average.


\section{Conclusions}
\label{sec:conclusions}
In this paper we have presented a non-parametric method to estimate the impulse response of a continuous-time system based on Lebesgue-sampled data. The proposed algorithm, which is inspired by MAP estimation and kernel methods, exploits the intersample knowledge of the output to deliver more accurate models than the standard approach while needing much fewer output samples. We have showed the efficacy of the method in a practically-relevant example. Future research in this setup includes the study of computationally-efficient algorithms and the allocation of the threshold levels for experiment design.

\bibliography{References}                                                            
\appendix
\section{Proof of Lemma \ref{lemma32}}
\label{prooflemma32}
\begin{pf}
	The MAP estimator for $\mathbf{c}$ is computed by
	\begin{align}
		\hat{\mathbf{c}}_{\textnormal{MAP}} \hspace{-0.09cm} &= \hspace{-0.07cm}\underset{\mathbf{c}\in \mathbb{R}^N}{\arg \max} \hspace{0.03cm}\big(\hspace{-0.03cm}\log \textnormal{p}(\mathbf{y}_{1:N}|\mathbf{c})+\log \textnormal{p}(\mathbf{c})\big) \notag \\
		&=\hspace{-0.07cm}\underset{\mathbf{c}\in \mathbb{R}^N}{\arg \max} \bigg(\hspace{-0.15cm}-\frac{N}{2}\hspace{-0.03cm}\log(\hspace{-0.02cm}2\pi \sigma^2\hspace{-0.02cm})\hspace{-0.07cm}-\hspace{-0.09cm}\frac{\log \det(\pi\mathbf{K}^{\hspace{-0.04cm}-\hspace{-0.02cm}1}/\gamma\hspace{-0.03cm})}{2}\notag \\
		\label{compareto}
		&\hspace{-0.02cm}+\hspace{-0.07cm} \sum_{i=1}^{N} \hspace{-0.02cm}\log\hspace{-0.07cm} \left[\hspace{-0.05cm}\int_{\eta_i}^{\eta_i\hspace{-0.03cm}+\hspace{-0.02cm}h}\hspace{-0.28cm} e^{\frac{-1}{2\sigma^2}\big(\hspace{-0.03cm}z_i\hspace{-0.03cm}-\hspace{-0.03cm}\mathbf{K}_i^\top \mathbf{c}\big)^{\hspace{-0.02cm}2}}\hspace{-0.09cm}\textnormal{d}z_i \hspace{-0.02cm}\right]\hspace{-0.13cm}-\hspace{-0.07cm}\gamma\mathbf{c}^{\hspace{-0.06cm}\top}\hspace{-0.03cm} \mathbf{Kc}\hspace{-0.06cm}\bigg),\hspace{-0.04cm}
	\end{align}
	where we have used the same derivation as for $\ell(g)$ in Theorem \ref{thm1} for computing the log-likelihood term. By comparing \eqref{compareto} to \eqref{computec}, we find that $\hat{\mathbf{c}}$ in \eqref{computec} is simply the MAP estimate of $\mathbf{c}$ within the model in \eqref{modelforc}. \hfill \hspace{2cm}\qed 
\end{pf}

\section{Proof of Theorem \ref{theorem31}}
\label{prooftheorem31}
\begin{pf}
Since the $Q$ function provided by Lemma \ref{lemma33} is concave in $\mathbf{c}$, it is sufficient to obtain the point(s) which make the gradient of the objective function equal to zero. The gradient of $Q(\mathbf{c},\hat{\mathbf{c}}^{(j)})-\gamma \mathbf{c}^\top \mathbf{Kc}$ is given~by
\begin{align}
	&\textstyle\frac{\partial}{\partial \mathbf{c}} \left( Q(\mathbf{c},\hat{\mathbf{c}}^{(j)})-\gamma\mathbf{c}^\top \mathbf{Kc}\right) \notag \\
	&= \hspace{-0.07cm} \frac{-1}{\sigma^2} \hspace{-0.05cm} \sum_{i=1}^{N} \hspace{-0.05cm} \int_{\eta_i}^{\eta_i\hspace{-0.02cm}+\hspace{-0.02cm}h}\hspace{-0.25cm} \mathbf{K}_i(\mathbf{K}_i^\top \hspace{-0.04cm} \mathbf{c}\hspace{-0.07cm}-\hspace{-0.07cm}z_i) \textnormal{p}(z_i|y(\hspace{-0.01cm}i\Delta\hspace{-0.01cm}),\hspace{-0.02cm}\hat{\mathbf{c}}^{(j)}\hspace{-0.02cm})\textnormal{d}z_i \hspace{-0.07cm}-\hspace{-0.07cm} 2\gamma \mathbf{Kc}\hspace{-0.02cm}. \notag
\end{align}
Setting the gradient to zero yields the iterations
\begin{equation}
	\label{intermediate}
	\hat{\mathbf{c}}^{(j+1)}\hspace{-0.05cm}=\hspace{-0.05cm} \left(\sum_{i=1}^{N} \mathbf{K}_i \mathbf{K}_i^\top \hspace{-0.05cm}+\hspace{-0.05cm} 2\gamma \sigma^2 \mathbf{K}\right)^{-1} \hspace{-0.1cm}\sum_{i=1}^{N} \mathbf{K}_i  \tilde{z}_i^{(j)},
\end{equation}
where we have defined the conditional mean $\tilde{z}_i^{(j)}$ as
\begin{equation}
	\label{intermediate0}
	\tilde{z}_i^{(j)} =  \int_{\eta_i}^{\eta_i+h} z_i \textnormal{p}(z_i|y(i\Delta),\hat{\mathbf{c}}^{(j)})\textnormal{d}z_i,
\end{equation}
and where we have taken into account that
\begin{equation}
	\int_{\eta_i}^{\eta_i+h} \hspace{-0.1cm}\textnormal{p}(z_i|y(i\Delta),\hat{\mathbf{c}}^{(j)})\textnormal{d}z_i = 1, \quad  i=1,2,\dots, N. \notag
\end{equation}
The iterations \eqref{ck1} are obtained from \eqref{intermediate} by rewriting the sum related to $\tilde{z}_i^{(j)}$ and using the fact that $\sum_{i=1}^{N} \mathbf{K}_i \mathbf{K}_i^\top = \mathbf{K}^2$, which holds since $\mathbf{K}$ is symmetric. Finally, the explicit expression for $\tilde{z}_i^{(j)}$ in \eqref{tildezi} can be obtained directly from expanding the following alternative expression for \eqref{tildezi} based on applying Bayes' theorem on the conditional expectation in \eqref{intermediate0}:
\begin{equation}
	\hspace{0.8cm}\tilde{z}_i^{(j)} = \frac{\int_{\eta_i}^{\eta_i+h}z_i \exp\left(\frac{-1}{2\sigma^2} [z_i-\mathbf{K}_i^\top \hat{\mathbf{c}}^{(j)} ]^2 \right) \textnormal{d}z_i}{\int_{\eta_i}^{\eta_i+h} \exp\left(\frac{-1}{2\sigma^2} [z_i-\mathbf{K}_i^\top \hat{\mathbf{c}}^{(j)} ]^2 \right) \textnormal{d}z_i}.  \notag \hspace{0.5cm}\hfill \qed 
\end{equation}
\end{pf}

\section{Proof of Theorem \ref{theorem32}}
\label{prooftheorem32}
	\begin{pf}
		We seek to derive the MAP-EM iterations for computing the estimator in \eqref{eb}. By setting the latent variable as $\mathbf{z}_{1:N}$, we must compute the following $Q$ function
		\begin{equation}
			Q(\bm{\rho},\hat{\bm{\rho}}^{(j)}) = \mathbb{E}\left\{\log \textnormal{p}(\mathbf{z}_{1:N},\mathbf{y}_{1:N}|\bm{\rho})|\mathbf{y}_{1:N},\hat{\bm{\rho}}^{(j)}\right\}. \notag
		\end{equation}
		It can be shown that (cf. Eq. (19) of \cite{godoy2011identification})
		\begin{equation}
			\textnormal{p}(\mathbf{z}_{1:N},\mathbf{y}_{1:N}|\bm{\rho}) = \begin{cases}
				\textnormal{p}(\mathbf{z}_{1:N}|\bm{\rho}) &\textnormal{if } \mathbf{z}_{1:N} \in \mathbf{y}_{1:N}, \\
				0 & \textnormal{otherwise,}
			\end{cases} \notag
		\end{equation}
		which, by exploiting \eqref{zrho}, leads to
		\begin{equation}
			-2Q(\hspace{-0.01cm}\bm{\rho},\hspace{-0.02cm}\hat{\bm{\rho}}^{\hspace{-0.01cm}(j)}\hspace{-0.01cm}) = \log \hspace{-0.02cm}\det(\hspace{-0.02cm}2\pi\mathbf{S}_{\hspace{-0.03cm}\bm{\rho}}\hspace{-0.02cm})\hspace{-0.03cm}+\hspace{-0.02cm}\mathbb{E}\{\hspace{-0.01cm}\mathbf{z}_{1\hspace{-0.01cm}:\hspace{-0.01cm}N}^\top \mathbf{S}_{\bm{\rho}}^{-\hspace{-0.02cm}1}\mathbf{z}_{1\hspace{-0.01cm}:\hspace{-0.01cm}N} \hspace{-0.01cm}|\mathbf{y}_{1\hspace{-0.01cm}:\hspace{-0.01cm}N}\hspace{-0.01cm},\hspace{-0.01cm}\hat{\bm{\rho}}^{(j)}\}. \notag
		\end{equation}
		The iterations in \eqref{emiterationsthm32} follow from applying the commutativity property of the trace to the expectation above.
		
		The minimization of $-2Q(\bm{\rho},\hat{\bm{\rho}}^{(j)})$ with respect to $\bm{\rho}$ provides the M-step of the MAP-EM iterations for computing a maximum of the likelihood of interest, which in turn is equivalent to solving the optimization problem in \eqref{kernelopt}. \qed
	\end{pf}
	
\end{document}